\documentclass[10pt,twocolumn,letterpaper]{article}

\usepackage{cvpr}              

\usepackage{graphicx}
\usepackage{amsmath}
\usepackage{amssymb}
\usepackage{booktabs}
\usepackage{multirow}
\usepackage{graphicx}
\usepackage[cmyk]{xcolor}
\usepackage{float}
\makeatletter
\renewcommand{\maketag@@@}[1]{\hbox{\m@th\normalsize\normalfont#1}}%
\makeatother

\usepackage[pagebackref,breaklinks,colorlinks]{hyperref}

\usepackage[capitalize]{cleveref}
\crefname{section}{Sec.}{Secs.}
\Crefname{section}{Section}{Sections}
\Crefname{table}{Table}{Tables}
\crefname{table}{Tab.}{Tabs.}

\def\cvprPaperID{*****} 
\def\confName{CVPR}
\def\confYear{2023}

\begin{document}

\title{
Team AcieLee: Technical Report for EPIC-SOUNDS Audio-Based Interaction Recognition Challenge 2023}

\author{Yuqi Li$^1$ \qquad  Yizhi Luo$^1$ \qquad Xiaoshuai Hao$^2$ \qquad Chuanguang Yang$^3$ \qquad Zhulin An$^3$\\
Dantong Song$^1$ \qquad Wei Yi$^4$\\ 
$^1$College of Computer and Information Science College of Software, Southwest University, Chongqing, China\\
$^2$Samsung Research China-Beijing(SRC-B)\\
 $^3$Institute of Computing Technology, Chinese Academy of Sciences, Beijing, China  \\ 
{\tt\small \{lllyq13, l3237614606, dtsong0420\}@email.swu.edu.cn}\\
{\tt\small \ xshuai.hao@samsung.com}\\
 {\tt\small \{yangchuanguang, anzhulin\}@ict.ac.cn}\\ 
 {\tt\small \ fire15@126.com}\\
}


\maketitle

\begin{abstract}

 In this report, we describe the technical details of our submission to the EPIC-SOUNDS Audio-Based Interaction Recognition Challenge 2023, by Team "AcieLee" (username: Yuqi\_Li). The task is to classify the audio caused by interactions between objects, or from events of the camera wearer. We conducted exhaustive experiments and found learning rate step decay, backbone frozen, label smoothing and focal loss contribute most to the performance improvement. After training, we combined multiple  models from different stages and integrated them into a single model by assigning fusion weights. This proposed method allowed us to achieve 3rd place in the CVPR 2023 workshop of EPIC-SOUNDS Audio-Based Interaction Recognition Challenge.


\end{abstract}

\section{Introduction}
\label{sec:intro}
With the rapid development of internet and social networks, multimodal data, e.g. image-text\cite{haoxiaoshuai4}, video-text\cite{haoxiaoshuai2023cvpr,haoxiaoshuai1,haoxiaoshuai2,haoxiaoshuai3,haoxiaoshuai5,Hao2020BidirectionalHR} and video-audio\cite{huh2023epic}, has attracted much attention. 
Sight and hearing are important perceptions that humans use to perceive the world. Audiovisual learning is a form of multimodal learning that uses both audio and video as input or output to achieve different tasks. The goal of audiovisual learning is to tap into the intrinsic connections between audio and video in order to improve the performance of single-modal tasks or solve new and challenging problems. The examples could be using lip movements in video to aid speech recognition, or using sounds in audio to generate corresponding images. The goal of the competition is to classify the audio caused by interactions between objects, or from events of the camera wearer.


\section{EPIC-SOUNDS Audio-Based Interaction Recognition Challenge}
\label{sec:intro}
EPIC-KITCHENS-100 is a dataset of unscripted egocentric visual actions collected from 45 kitchens in 4 cities worldwide. EPIC-Sounds expands on this dataset by annotating audio-based interactions that occur in the 100 hours of untrimmed video footage~\cite{codalab_competitions}.
EPIC-SOUNDS is a large-scale dataset of audio annotations that captures the temporal boundaries and class labels within the audio stream of egocentric videos. 

The dataset conducts an annotation process where annotators label distinguishable audio segments over time and describe the action that could generate sound. By grouping free-form audio descriptions into classes, actions that can be differentiated solely based on audio are identified. For actions that involve the collision of objects, human annotations of the materials of these objects (e.g., a glass object being placed on a wooden surface) are gathered and confirmed using visual labels and eliminating any ambiguities. In total, EPIC-SOUNDS comprises 78.4k categorized segments of audible events and actions spreading across 44 classes, as well as 39.2k non-categorized segments~\cite{Damen2022RESCALING,huh2023epic}. The goal of this competition is to classify the audio resulting from interactions between objects or from events involving the camera wearer. 

The dataset for this competition can be found on GitHub at https://github.com/epic-kitchens/epic-sounds-annotations. The EPIC-SOUNDS dataset is split into two parts: the media data (audio files) and the training annotations~\cite{codalab_competitions}. And the evaluation criteria are: top-1 and top-5 accuracy, mean average precision (mAP), mean Area Under Curve (mAUC), mean Per-Class Accuracy (mCA) on all instances in this set.
\begin{figure*}[htbp]
\centering
\includegraphics[totalheight=2in]{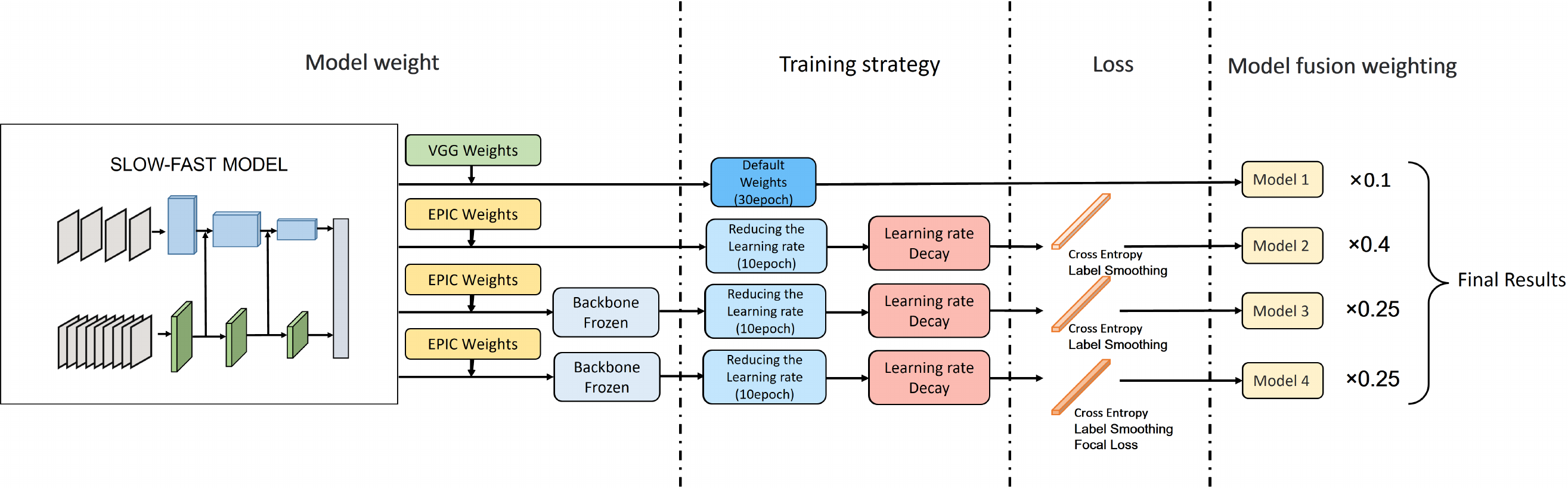}
\caption{The pipeline of our method. We constructed four models and then pre-trained them, froze the backbone, reduced the number of epochs, and added a cross-entropy loss function with label smoothing and focal loss. Finally, we assigned fusion weights to the four trained models to obtain the final result.}
\end{figure*}

\section{Proposed method}
\label{sec:formatting}
\subsection{Training Pipeline}
In the model weights section, we use the Slow-Fast algorithm model to construct four training network models. After that, the weights of the VGG pre-training model and the EPIC pre-training model are loaded in the Slow-Fast model for training respectively~\cite{xiong2022m}. The third and fourth EPIC pre-trained models are frozen during training to speed up the training efficiency and prevent weight destruction.

During the frozen stage, the backbone of the model is frozen, and the feature extraction network remains unchanged, occupying less GPU memory. Only fine-tuning is performed on the network. During the unfrozen stage, the backbone of the model is unfrozen, and the feature extraction network underwent changes. All parameters of the network are modified.

In the training phase, for the first VGG pre-trained model, it is trained for 30 epochs using the default parameters. For the three EPIC pre-trained models, they are trained for 10 epochs with a reduced learning rate using  a step-wise decay

Next, regarding the loss function, the first and second EPIC pre-trained models adopt the cross-entropy loss function with label smoothing. For the fourth EPIC pre-trained model, in addition to using the cross-entropy loss function and label smoothing~\cite{muller2019does}, we also use focal loss~\cite{lin2017focal}.

Finally, weights are assigned to the four trained models, in the order of 0.1, 0.4, 0.25, and 0.25, respectively, for the final submission of results.  Our pipeline is shown in Fig 1.

\subsection{Loss Function}
Our baseline uses the original cross-entropy loss function:

\begin{equation}
\mathcal{L} _{CE}\left( y,p \right) =
\begin{cases}
	-\log \left( p_i \right) ,   \qquad \quad   \left( i=c \right)\\
	-\log \left( 1-p_i \right) , \quad \ \left( i\ne c \right) \,\,\\
\end{cases}
\end{equation}

Where $C$ represents the number of categories, $c$ represents the category corresponding to the data, $y$ represents the label, which is in one-hot encoding form. $p=[p_0, p_1,..., p_{C-1}]$ is a probability distribution, with each element $p_i$ representing the probability that the sample belongs to the $i$-th class. $y=[y_0, y_1,..., y_{C-1}]$ is the one-hot representation of the sample label, where $y_i=1$ when the sample belongs to class $i$ and $y_i=0$ otherwise. Which means:

\begin{equation}
y_i=\begin{cases}
	1,   \quad \left( i=c \right)\\
	0,   \quad \left( i\ne c \right)\\
\end{cases}
\end{equation}

After introducing label smoothing, the loss function becomes:

\begin{equation}
y_i=\begin{cases}
	\left( 1-\varepsilon \right) , \quad \left( i=c \right)\\
	\frac{\varepsilon}{C-1}, \qquad \  \left( i\ne c \right) \,\,\\
\end{cases}
\end{equation}

As mentioned before, $C$ represents the number of classes and $c$ represents the class to which the data corresponds. The loss function becomes:
\begin{equation}
L\left( y,p \right) =\begin{cases}
	-\left( 1-\varepsilon \right) *\log \left( p_i \right) ,   \quad  \   \ \left( i=c \right)\\
	-\frac{\varepsilon}{C-1}*\log \left( 1-p_i \right) ,  \quad \left( i\ne c \right)\\
\end{cases}
\end{equation}

After introducing gamma from the focal loss to make the model pay more attention to difficult samples and prevent overfitting, the loss function becomes:

\begin{equation}
L\left( y,p \right) =\begin{cases}
	-\left( 1-p \right) ^{\gamma}*\left( 1-\varepsilon \right) *\log \left( p_i \right) ,   \  \left( i=c \right)\\
	-p^{\gamma}*\frac{\varepsilon}{C-1}*\log \left( 1-p_i \right) ,     \qquad   \ \left( i\ne c \right)\\
\end{cases}
\end{equation}


\begin{table*}[!t]
\caption{Audio-Based Interaction Recognition results on EPIC-SOUNDS TEST sets}
\resizebox{\textwidth}{!}{
\begin{tabular}{|c|c|c|c|c|c|c|}
\hline
Rank           &Users                  & Top-1 Accuracy(\%)                                      & Top-5 Accuracy(\%) & Pre-Class Accuracy(\%) & Mean Average Precision(\%) & Mean Area Under Curve                                    \\ \hline
1              &stevenlau                & 55.43 (1)                                               & 85.52 (2)          & 21.84 (2)              & 26.98 (1)                  & 0.877 (1)
                        \\ \hline
2              &audi666                  & 55.11 (2)                                               & 85.40 (3)          & 21.14 (3)              & 25.96 (4)                  & 0.856 (3)                                               \\ \hline
\textbf{3}              &\textbf{ Yuqi\_Li (OURS)}                     & \textbf{55.07 (3)}                                               & \textbf{85.61 (1)}          & \textbf{20.95 (4)}              & \textbf{26.20 (2)}                  & \textbf{0.859 (2)}                                                \\ \hline
\textcolor{lightgray}4              &\textcolor{lightgray}{EPIC\_AUDITORY\_SLOWFAST} & \textcolor{lightgray}{54.80 (4)}                                               & \textcolor{lightgray}{85.18 (4)}          & \textcolor{lightgray}{20.77 (5)}              & \textcolor{lightgray}{26.01 (3)}                  & \textcolor{lightgray}{0.850 (4)}                                                \\ \hline
\textcolor{lightgray}5             &\textcolor{lightgray}{EPIC\_SSAST}              & \textcolor{lightgray}{53.97 (5)}                                               & \textcolor{lightgray}{84.53 (6)}          &\textcolor{lightgray} {22.41 (1)}              & \textcolor{lightgray}{25.43 (5)}                  & \textcolor{lightgray}{0.819 (6)}                                                 \\ \hline
\end{tabular}}
\end{table*}

\section{Experiment}
\subsection{Datasets}
\textbf{EPIC-SOUNDS.}
We use EPIC-SOUNDS data, a large scale dataset of audio annotations capturing temporal extents and class labels within the audio stream of the egocentric videos from EPIC-KITCHENS-100. EPIC-SOUNDS includes 78.4k categorised and 39.2k non-categorized segments of audible events and actions, distributed across 44 classes. The labeled temporal timestamps are provided for the train/val split, and just the unlabeled timestamps are utilized for the recognition test split. The temporal timestamps for annotations that could not be clustered into one of the 44 classes are also provided, along with the free-form description used during the initial annotation. Two state-of-the-art audio recognition models are trained and evaluated on this dataset, for which the code and pre-trained models are also provided~\cite{Damen2022RESCALING}.

\subsection{Evaluation metrics}
In this competition, the evaluation will be conducted from five aspects: top-1 and top-5 accuracy, mean average precision (mAP), mean Area Under Curve (mAUC), and mean Per-Class Accuracy (mCA). Our final submission ranked 3rd in Top-1 Accuracy, 1st in Top-5 Accuracy, 4th in Per-Class Accuracy, 2nd in Mean Average Precision, and 2nd in Mean Area Under Curve. In the end, our submission ranked third overall.

\subsection{Implementation Details}

\textbf{Feature extraction.}
We reduced the learning rate from 0.001 to 0.0001 for slight fine-tuning because we used a pre-trained model. A lower learning rate means that the speed of parameter update will slow down, which helps to adjust the model parameters more finely during the fine-tuning process. The total number of training epochs was reduced from 30 to 10 to shorten the time cost.

\textbf{Train / Val details.}
Step decay of the learning rate can make it easier for the model to find the minimum value. We set the step decay of the learning rate to [0, 2, 4, 6, 8], [1, 0.7, 0.5, 0.3, 0.1]. This means that the base learning rate is multiplied by 1, 0.7, 0.5, 0.3, and 0.1 at 0, 2, 4, 6, and 8 epochs, respectively~\cite{ge2019step}. At the same time, considering that there may have errors in data labeling, the original one-hot encoding is too absolute. To reduce overfitting and improve generalization ability, we replaced the original default cross-entropy and added label-smoothing with a value of 0.06~\cite{muller2019does}.

After these operations, we obtained a local validation score of 53.71. After submission, the online test result was 54.94, surpassing EPIC\_AUDITORY\_SLOWFAST.
Next, we set FREEZE\_BACKBONE as True to freeze the backbone and tested it on the local validation set. Freezing the backbone is a common trick. Once the model is almost trained, the backbone is frozen and the head part is fine-tuned to improve the model accuracy. We obtained a score of 53.79. Since the improvement was small and there is a limit of two submissions per day, we did not submit it for online testing.

Then, in order to make the model pay more attention to difficult samples and prevent overfitting, we added gamma to the cross-entropy loss with focal loss, setting it to 0.3~\cite{lin2017focal}. This time we obtained the highest local validation score of 53.83, but the online test score was 54.48. Therefore, we chose to merge the previous experimental results and perform weighting~\cite{byrd2019effect}, as shown in Table 2. Because the score of the first model is low and the second model has the highest score in the online test, they are given smaller (empirically chosen 0.1) and larger (empirically chosen 0.4) weights respectively. The third model was not submitted for online testing due to submission limits. However, its score on the offline validation set is very close to that of the fourth model. The fourth model is close to the second model but slightly lower than the second model in online testing. Therefore, we set 0.25 for them based on expert experience. In the end, we obtained a merged score of 55.07 for the online test, which is also our final result.
\begin{table}[]
\caption{Fusion experiment and weight assignment}
\centering
\resizebox{\linewidth}{!}{%
\begin{tabular}{ccccc}
\hline
\multirow{2}{*}{No.} & \multirow{2}{*}{Experiment} & \multicolumn{2}{c}{Results}                           & \multirow{2}{*}{weighting} \\ \cline{3-4}
  &     & Val   & \multicolumn{1}{c}{Test} &      \\ \hline
1                    & Baseline                    & \multicolumn{1}{r}{50.53} & \multicolumn{1}{r}{52.31}       & 0.1                      \\
2 & 1+LR step decay & 53.71 & 54.94                       & 0.4  \\
3 & 2+freeze backbone & 53.79 & No                         & 0.25 \\
4 & 3+gamma:0.3 & 53.83 & 54.58                       & 0.25 \\ \hline
\end{tabular}%
}
\end{table}
\subsection{Main Results}
 According to the experimental details section, by lowering the learning rate, reducing the number of epochs, setting the learning rate to step-wise decay, replacing the original default cross-entropy, adding label smoothing, freezing the backbone, adding gamma from focal loss to the cross-entropy loss function, and then weighting and combining the experimental results from these different processes, we obtained the final combined score and highest online test score of 55.07.

\subsection{Experimental protocol}
We attempted to reproduce the official results and obtained a local validation score of 50.53 and an online test score of 52.31. We used this result as our baseline. By following this setting, we loaded the pre-trained VGG model and used the official provided config with 30 training epochs~\cite{panayotov2015librispeech,chen2020vggsound}. Next, we loaded the official epic model as a pre-trained model and attempted to make improvements.

\section{Conclusion}
In this competition, we fused four models together and assigned weights. We demonstrated the importance of our proposed model on the EPIC-SOUNDS dataset and achieved SOTA performance. In addition to the effective methods mentioned above, we also tried many new methods during the participation process, for example, reducing the batch size, adjusting the dropout~\cite{srivastava2014dropout} ratio, adding drop-path~\cite{larsson2016fractalnet}, min-max normalization preprocessing, mixup~\cite{zhang2017mixup}, cutmix~\cite{yun2019cutmix}, and spectral data augmentation (random occlusion, brightness, noise, etc). However, these practices cannot improve results.

For future work, we believe that using more powerful backbones, such as HCGNet~\cite{yang2020gated}, ResNeSt~\cite{zhang2022resnest}, and Vision Transformer~\cite{dosovitskiy2020image}, may yield better results. Moreover, Knowledge Distillation (KD)~\cite{hinton2015distilling} is also a common paradigm to optimize the model. We think using our previously proposed KD methods, such as HSAKD~\cite{yang2021hierarchical}, CIRKD~\cite{yang2022cross}, L-MCL~\cite{yang2022mutual,yang2023online} and MixSKD~\cite{yang2022mixskd}, could enhance the model accuracy. We will also try these techniques after the competition.
     

{\small
\bibliographystyle{ieee_fullname}
\bibliography{ref}
}
\end{document}